\documentclass[aps,pra,epsf,preprint,superscriptaddress,showpacs,preprintnumbers,amsmath,amssymb,11pt]{revtex4-1}
\usepackage{graphicx}
\usepackage{epsfig}
\usepackage{dcolumn}
\usepackage{bm}
\usepackage{braket}
\usepackage{color}
\usepackage{amsmath}
\begin{document}

\title{Structure of the correlation function at the accumulation\\
points of the logistic map}

\author{K. Karamanos}
\affiliation{Complex Systems Group, Institute of Nuclear and
Particle Physics, NCSR Demokritos, GR 15310, Aghia Paraskevi Attiki,
Greece}
\author{I.S. Mistakidis}
\affiliation{Hellenic Army Academy, Section of Physical Sciences and
Applications, Vari Attiki, Greece}
\author{S.I. Mistakidis}
\affiliation{Zentrum f\"{u}r Optische Quantentechnologien,
Universit\"{a}t Hamburg, Luruper Chaussee 149, 22761 Hamburg,
Germany}

\date{\today}

\begin{abstract}

The correlation function of the trajectory exactly at the Feigenbaum
point of the logistic map is investigated and checked by numerical
experiments. Taking advantage of recent closed analytical results on
the symbol-to-symbol correlation function of the generating
partition, we are in position to justify the deep algorithmic
structure of the correlation function apart from numerical
constants. A generalization is given for arbitrary $m
\cdot 2^{\infty}$ Feigenbaum attractors.\\

\end{abstract}

 \maketitle

\section{Introduction}

Recently, the study of Complex Systems has gained significant
attention. One of the basic aspects of this progress is related with
the understanding of correlations in and between such complex
systems, which is realized through the use of different complexity
measures. Among these, one can mention the transinformation
\cite{NicolisG,Nicol,Nicolis91}, the block entropies
\cite{Athan,Grassb86,Karamano00,Karam1,
Karam01,Karam02,Karam03,Karam99} different types of correlation
functions \cite{Alon96,Bai94,Crutchfield1,May,Schroder,Schuster} 
and number-theoretic notions \cite{Karamano00,Karaman01}.

One of the Paradigms of Complex Systems is the logistic map. The
logistic map has a simple definition but presents complex behavior
when fine tuning the control parameter values. In particular, after
Feigenbaum's work, the period-doubling route to chaos has been
fairly understood. Also, connections with the theory of second order
phase transitions (critical phenomena) have been established and
scaling relations have been reported nearby the accumulation point
(also called Feigenbaum Point (FP)) with and without the presence of
external noise. Furthermore, cantorian fractal structures have been
revealed in the transition point connecting the physics of the
non-chaotic attractor with self-similarity
\cite{Crutchfield,Ebel91,Ebel92,Feige,Feigen91,Feigen82,Freud96}.
Recently also, a direct connection with Experimental Mathematics has
been established, too \cite{Karam100}.

On the other hand, in Non-linear physics, the importance of the
study of the correlation function has been realized from the very
beginning. Particularly inspiring have been the works of Ruelle
\cite{Ruelle}, Daems and Nicolis \cite{Daems94}, and Alonso et al.
\cite{Alon96}, for the case of resonances of chaotic dynamical
systems. In addition, based on the analogies between the period
doubling transition and critical phenomena, H.~G. Schuster has done
a guess on the functional form of the correlation function of the
trajectory \cite{Schuster}. Indeed, according to his arguments the
correlation function should follow a power law behaviour. 
In contrast, here, we demonstrate that the correlation 
function possesses a stratified structure.
More recently, using the Feigenbaum renormalization group transformation it has been shown \cite{Robledo} 
that the correlation function of the trajectory in the one dimensional nonlinear dissipative 
logistic map is made of a family of power laws with a common scaling factor given by the 
Feigenbaum constant $\alpha$. 
In the present work in order to extract the form of the correlation function of the trajectory we propose 
some more elaborated arguments, using a different approach which is based  
on the structure of the symbol-to-symbol correlation function
\cite{Karam03}, that is the correlation function of symbolic
dynamics.

 After establishing rigorously in a previous work \cite{Karam03} the detailed form of the
symbol-to-symbol correlation function we turn now our attention to
the structure of the correlation function of the trajectory. To be
more concrete, taking advantage from the analytic form of the
symbol-to symbol correlation function and presenting simple
arguments we shall show that one can extract up to a good
approximation, that is apart from numerical constants, the detailed
structure for the correlation function of the trajectory. The above
investigation is mainly supported by a detailed numerical study
which takes into account a large enough statistical sample of the logistic map.
In this manner, we can justify the analytic form of the correlation function of
the trajectory from first principles using the Metropolis-Stein and Stein algorithm (MSS algorithm), apart from
numerical constants, which depend on the detailed functional form of
the map. Furthermore, we make an attempt to generalize these results for an 
arbitrary $m \cdot {2^\infty }$ accumulation point \cite{Grassberger}, for $m=2,3,...$, which correspond 
to the accumulation points of the bifurcation tree \cite{Schuster,Luo} (see also Figure 1). Finally, a general form 
for the correlation function of the trajectory and that obtained from the 
symbolic dynamics is also suggested.  
We believe that our results will inspire similar investigations on
non-unimodal maps and give further insight providing new complexity
measures on real experimental time-series.

The paper is organized as follows. In Sec. II we introduce the logistic map and the definitions
of different types of correlation functions that will be used. In
Sec. III we present our careful numerical experimentation for the
symbol-to-symbol correlation function and for the correlation
function of the trajectory at the (first) accumulation point. As it is 
shown those functions satisfy simple numerical prescriptions, which
are explicitly outlined. In addition, we propose some simple
arguments which, up to a good approximation, allow for the
justification of the functional form of the correlation function of
the trajectory from the symbol-to-symbol correlation function apart
from arithmetical constants in a systematic basis. We then present
analogous results and generalizations for the $m \cdot {2^\infty }$
accumulation points. Finally, in Sec. IV we draw the main
conclusions and discuss future plans.

\section{The logistic map}
The logistic map is the archetype of a Complex System. Let us
elaborate. We introduce the logistic map in its familiar form
\begin{equation}
\label{eq:2}{x_{n + 1}} = r{x_n}(1 - {x_n}),
\end{equation}
where $r$ is the control parameter value and $n$ denotes the
respective iteration of the map. For the logistic map in this form
the generating partition is easily computed, following an argument
dating back to the French Mathematician Gaston Julia. To be more
specific, for $f(x) = rx(1 - x)$ the equation $f'(c) = 0$ gives
c=0.5, so that the partition of the phase space (which in this case
coincides with the unit interval I=[0,1]) $L$=[0,0.5] and $R$=(0.5,1] is
a generating one (see also \cite{Beck} for a more rigorous definition). Notice that according to Metropolis et al.
\cite{Metropolis} the information content of the symbolic trajectory
is the "minimum distinguishing information". Needless to say, in
this representation the logistic map can be viewed as an abstract
information generator.

\begin{figure}[ht]
\vspace{-10pt}
        \centering
                \includegraphics[width=0.80\textwidth]{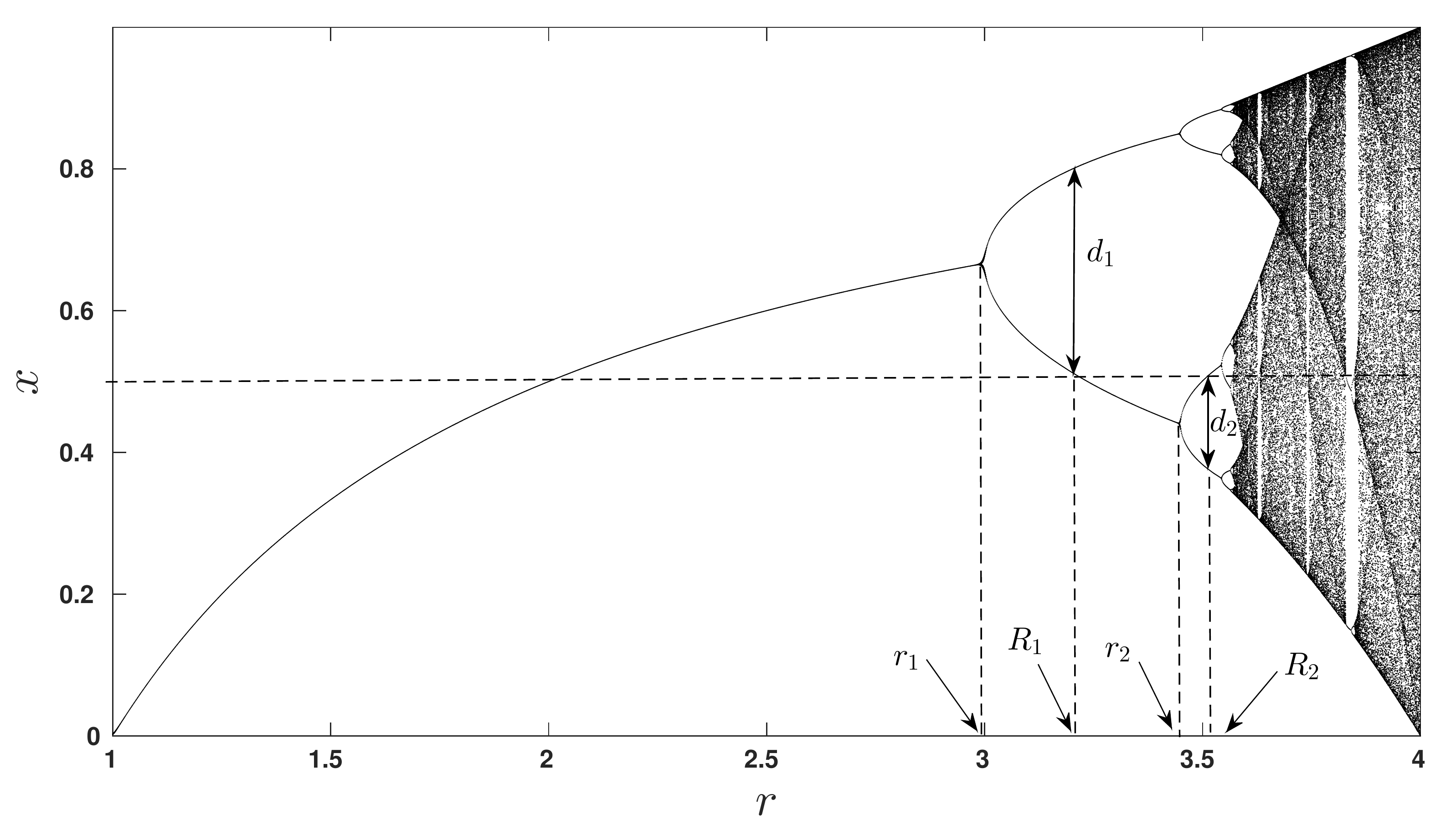}
                \vspace{0.0cm}
                \caption{The bifurcation diagram for the logistic map for the superstable ${2^n}$-cycles. It is shown
                the control parameter values $r_{i}$ for the first few bifurcation points and the values $R_{i}$ for
                the superstable orbits.}
\end{figure}
In particular, the period doubling route to chaos has been fairly
studied and it is by now well understood. These studies led to
the occurrence of the two Feigenbaum constants $\alpha $ and $\delta
$ which can be defined by an approximate real space renormalization
procedure. Especially, the constant $\delta$ is related with the
spacing in the control parameter space of the successive values of
occurrence of the superstable periodic orbits and can be roughly
estimated by the bifurcation diagram \cite{Feige,Feigen91}. If we
denote as $\{ {R_n}\}$ this set of values, $\delta$ is defined as
\begin{equation}
\label{eq:3}\delta  = \mathop {\lim }\limits_{n \to \infty }
\frac{{{R_n} - {R_{n - 1}}}}{{{R_{n + 1}} - {R_n}}},
\end{equation}
and for the quadratic map reads
\begin{equation}
\label{eq:4}\delta  \simeq 4.669201609102990... ~~.
\end{equation}
Moreover, the constant $\alpha$ is related to the rescaling of the
period doubling functional composition law and its value for the
logistic map reads
\begin{equation}
\label{eq:5}\alpha= - \mathop {\lim }\limits_{n \to \infty }
\frac{{{d_n}}}{{{d_{n + 1}}}}  \simeq -2.5029078750095892...~~.
\end{equation}
Finally, note that the constants $\alpha$, $\delta$ are related as it can be shown    
by using renormalization group arguments (see \cite{Karamanos10,Schroder} and references therein). 
The values of the above two constants depend only
on the order of the maximum and have long been studied. They are
thus, for instance, universal for quadratic maps irrespectively of
the exact way one writes down the map.

Figure 1 presents the control parameter values of the bifurcation
points denoted as ${r_1}$, ${r_2}$, ${r_3}$,... while the
corresponding values for the superstable orbits are depicted as
${R_1}$, ${R_2}$, ${R_3}$,... . The values of ${d_i}$ figuring in
the definition of the Feigenbaum constant $\alpha$ are also shown.
Note here that Feigenbaum and successors have shown that eq.(2),
holds if instead of ${R_i}$ we use ${r_i}$.

After the above brief introduction of the logistic map and its
properties, we shall next define the (un-normalized) correlation
function of the trajectory as
\begin{equation}
\label{eq:7}{C_{un}(m) = \mathop {\lim }\limits_{N \to \infty }
\frac{1}{N}\sum\limits_{i = 0}^{N - 1} {{{x}_{i +
m}}{{x}_i}}, \quad }
\end{equation}
where the deviation from the real value of the map at the $i-th$
iteration is given by ${{{x}_i} = {f^i}({x_0}) - \overline
x }$ and the corresponding mean value of the map taking into account
$N$ iterations (sample) is denoted by ${\quad {\overline x }=
\mathop {\lim }\limits_{N \to \infty } \frac{1}{N}\sum\limits_{i =
0}^{N - 1} {{f^i}({x_0})}}$. Also, in direct analogy with the above defined 
un-normalized correlation function one can also introduce here the
normalized correlation function
\begin{equation}
\label{eq:8}C(m) = \frac{{{C_{un}}(m)}}{{{C_{un}}(0)}} =
\frac{{{C_{un}}(m)}}{{{\sigma ^2}}},
\end{equation}
where $\sigma $ is the mean standard deviation, which normalizes the
statistical data.

From the above definitions follows that $C(m)$ (or equally
$C_{un}(m)$) yields another measure for the irregularity of the
sequence of iterates ${x_0}$, $f({x_0})$, ${f^2}({x_0})$,...etc. It
tells us how much the deviations of the iterates from their average
value, ${x_i} = {x_i} - \overline x$ that are m steps apart
(i.e. ${x_{i + m}}$ and ${x_i}$) "know" about each
other, on the average. Another remark here is that if
$C(m)\nrightarrow 0$ as $m \to \infty $ then the system does not
have the mixing property.

We should here note that the problem of determining the correlation
function of an arbitrary dynamical system is difficult to calculate in the
general case. This is the reason to resort to other computable
observables such as the symbol-to-symbol correlation function
\cite{Daems94}. Thus, in direct analogy with the correlation
function of the trajectory one can introduce the un-normalized
symbol-to-symbol correlation function as
\begin{equation}
\label{eq:9}{K_{un}(m) = \mathop {\lim }\limits_{N \to \infty }
\frac{1}{N}\sum\limits_{i = 0}^{N - 1} {{{y}_{i +
m}}{{y}_i}}},
\end{equation}
where the corresponding symbolic functional composition is defined by
${{{{ y}_i} = y({f^i}({x_0})) - \overline y }}$, with the
mean value ${{\overline y  = }\mathop {\lim }\limits_{N \to \infty }
\frac{1}{N}\sum\limits_{i = 0}^{N - 1} {y({f^i}({x_0}))}}$. Here,
$f^{i}(x_0)$ is the real value of the map at the $i-th$ iteration,
$N$ denotes the size of the statistical sample that we take into
account and $m$ is the corresponding distance between two symbols in
the symbolic sequence that we examine. In addition, as usual the
functional composition ${y_i}$ takes the values ${
y_i}$=0,1 when ${x_i} \leqslant 0.5$ or ${x_i}> 0.5$ respectively, i.e. it is a step function.

\begin{figure}[ht]
\vspace{-15pt}
        \centering
                \includegraphics[width=0.80\textwidth]{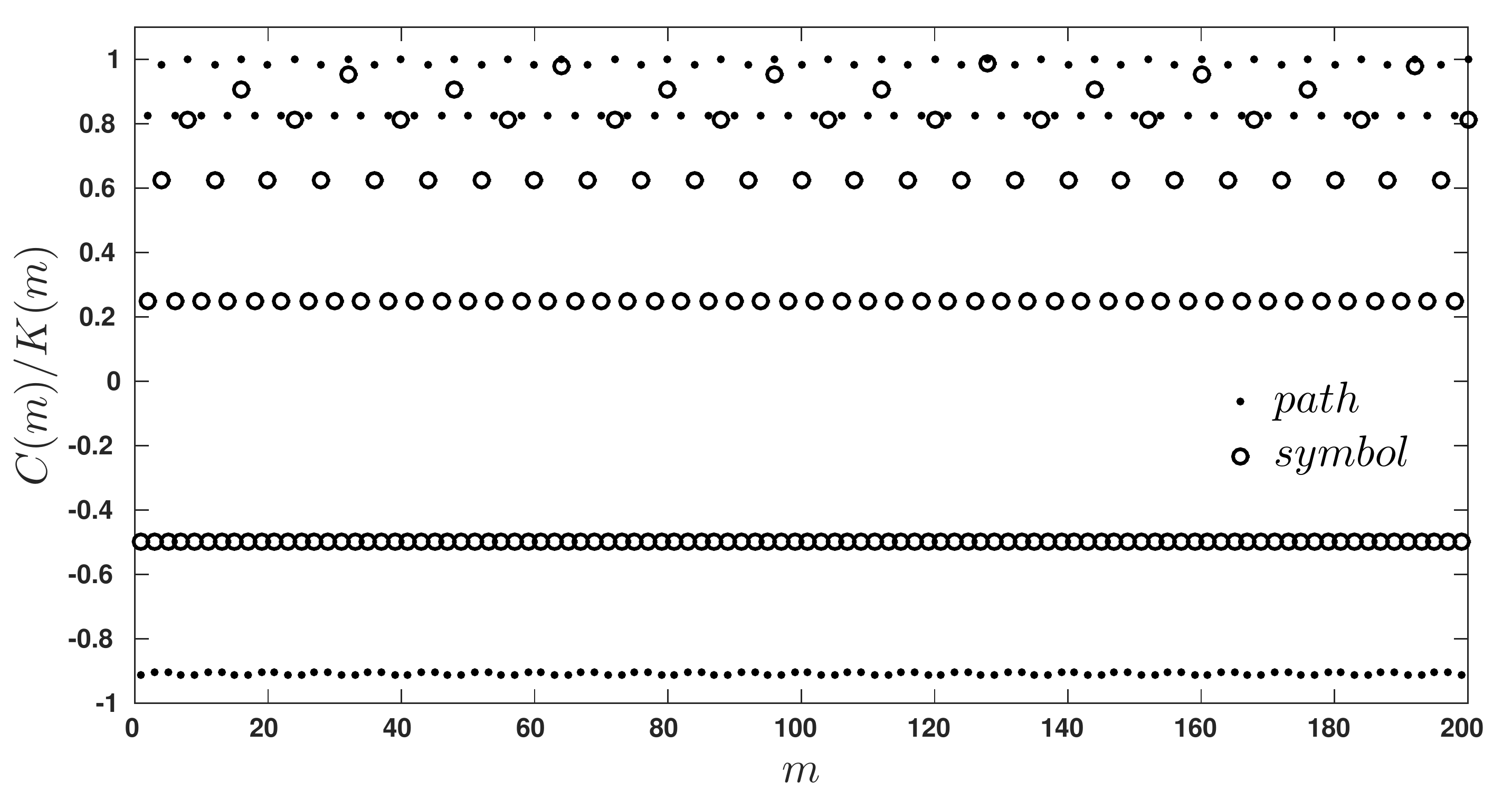}
                \vspace{-0.2cm}
                \caption{Shown are the symbol-to-symbol correlation function and the correlation function of
                the trajectory
                for the logistic map exactly at the accumulation point $r =
                FP = 3.56994567...$, with initial condition ${x_0} = 0.5$. The first ${10^5}$ iterations
                have been eliminated from our statistics in order to
                exclude transients, and the subsequent $n={10^8}$ iterations have been taken into
                account for the numerical calculations. The experimentally determined Lyapunov exponent
                 is $ - 5.93 \cdot {10^{ - 5}}$.}
\end{figure}
Finally, let us also note that one can define a more relevant
quantity which is the normalized symbol-to-symbol correlation
function
\begin{equation}
\label{eq:11}{\rm K}(m) = \frac{{{{\rm K}_{un}}(m)}}{{{{\rm
K}_{un}}(0)}} = \frac{{{{\rm K}_{un}}(m)}}{{{{\sigma '}^2}}},
\end{equation}
where $\sigma' $ denotes the mean standard deviation obtained from
the respective symbolic sequence. It has been shown \cite{Wentian,Herzel} that the normalized 
symbol-to-symbol correlation function does not depend on the choice of 
the symbols (i.e. $L$ to ''0'' and $R$ to ''1'') because for binary sequences 
the correlation functions are unique up to a constant factor, which varies with 
the choice of these numbers but cancels out via the normalization.

\section{The structure of the correlation function}

Motivated by previous works on correlation functions
\cite{Daems94,Ruelle,Robledo}, we explore here the properties of the
symbol-to-symbol correlation function and the correlation function
of the trajectory. In order to cope with the problem of the analytic
form of correlation functions we have studied the correlation function numerically.  
For the logistic map at the Feigenbaum point,
i.e. $r=3.56994567...$, we have calculated both the normalized
symbol-to-symbol correlation function (see eq.(8) and Figure 2) and
the correlation function of the trajectory (see eq.(6) and Figure
2). To do that we start each time from the initial point ${x_0} =
0.5$ and take a numerical sample consisting of $n={10^8}$ iterations
after the elimination of the first ${10^5}$ iterations (to avoid
transients). For reasons of completeness let us note that we have checked 
that our results, presented below, pertain if we use a different initial condition in 
the unit interval e.g. $x_0=0.3$ or $x_0=0.8$. 
Also we remark that according to our simulations (omitted here for brevity) the structure of the correlation function (see below) remains the same if instead   
of a single initial value (e.g. $x_0=0.5,0.3$) we average the 
correlation function over a uniform ensemble of initial values $x_0 \in \{0.0225,0.975\}$ with step $\delta x_0=0.0225$.
At this point we should remind that exactly at the
Feigenbaum point the Lyapunov exponent, which is defined in general
as $\lambda=(1/n)\lim_{n \to \infty} \sum\limits_{i=1}^n
{ln|f'(x_{i})|}$, strictly vanishes i.e. $\lambda= 0$ (see also
Table I), and we are in the presence of the non-chaotic multifractal
attractor. Notice also that the same behaviour of the Lyapunov
exponent holds for the higher accumulation points. To indicate the
behaviour of the map Table I presents the numerically calculated
Lyapunov exponent for every accumulation point (using eight decimals for the corresponding control parameter) 
of the logistic map including and excluding transients from our statistics. We observe
that the Lyapunov exponent in each case vanishes, while the
transients play no essential role due to the augmented statistics
that we use.

\begin{figure}[ht]
\vspace{-15pt}
        \centering
                \includegraphics[width=0.80\textwidth]{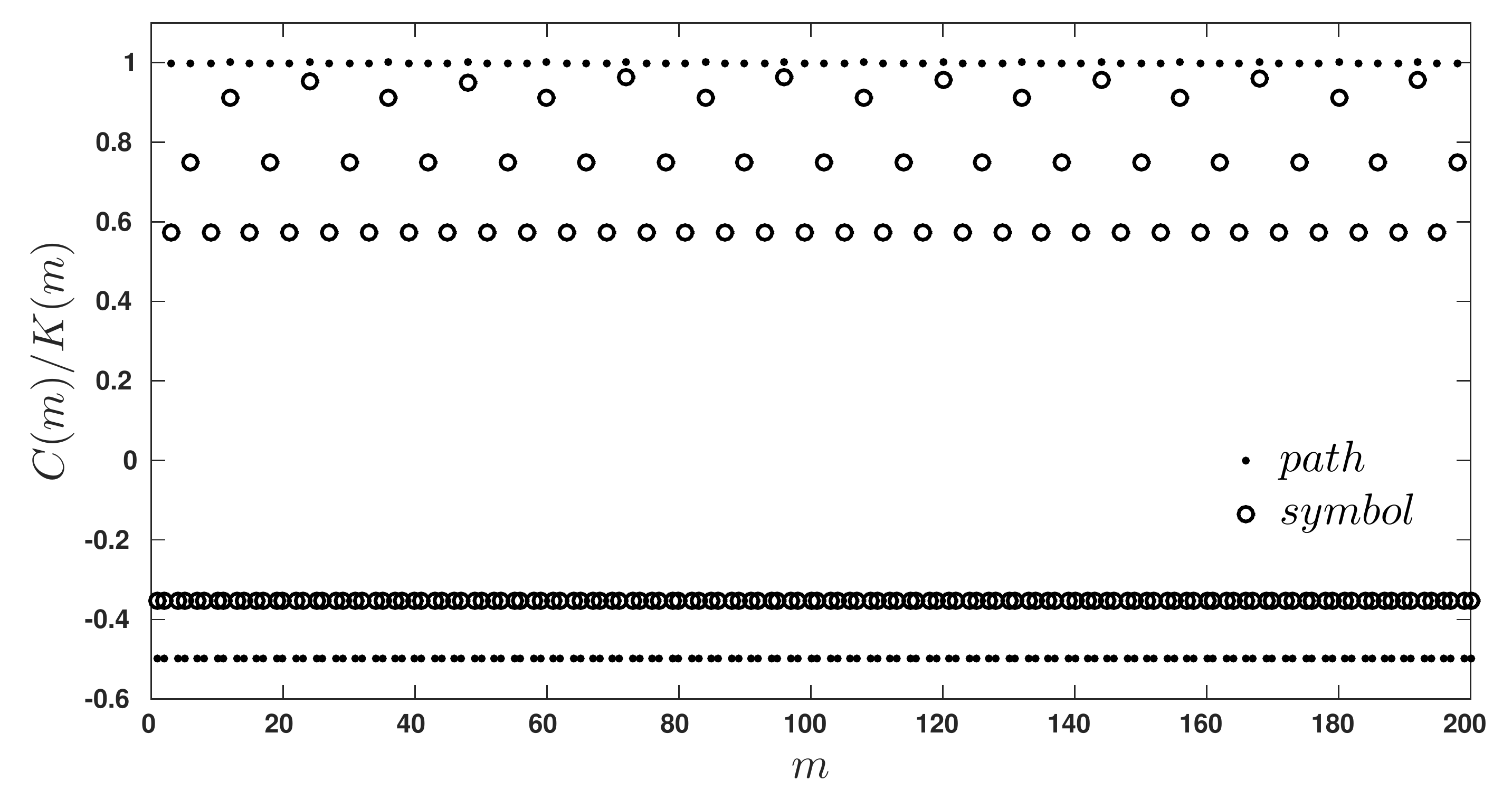}
                \vspace{-0.2cm}
                \caption{Shown are the symbol-to-symbol correlation function and the correlation function of the
                trajectory
                for the logistic map with control parameter value $r =3.8495$ (cycle $3\cdot{2^\infty }$) and initial condition
                 ${x_0} = 0.5$. The first ${10^5}$ iterations
                have been eliminated from our statistics in order to exclude transients, and the subsequent $n={10^8}$ iterations have been taken into
                account for the numerical calculations.}
\end{figure}
On the other hand, as it has already been mentioned, in a previous work
\cite{Karam03} providing some theoretical arguments from the
viewpoint of the symbolic dynamics we have established the structure
of the un-normalized symbol-to-symbol correlation function. These
results have also been supported from careful numerical
experimentations leading to the compact form
\begin{equation}
\label{eq:12}{K_{un}(m) = {A_l} \cdot {\delta _{m,{2^{l - 1}} \cdot
(1 + 2 \cdot k)}}},
\end{equation}
where for a given (fixed) $l$, $l \in \{ 1,2,3,....\}$, ${A_l}$ is a
constant depending only on $l$, and $k$ takes all the values from
the set of natural numbers $\{ 0,1,2,3,....\}$. Here, we have also
used the fact that any integer $m$ can be decomposed in terms of a
unique pair of natural numbers $l$, $k$ such that $m=2^{l-1}(1+2k)$.
The same holds for any other form of the correlation function that
will be presented in the rest of the paper. 
\begin{figure}[ht]
\vspace{-12pt}
        \centering
                \includegraphics[width=0.80\textwidth]{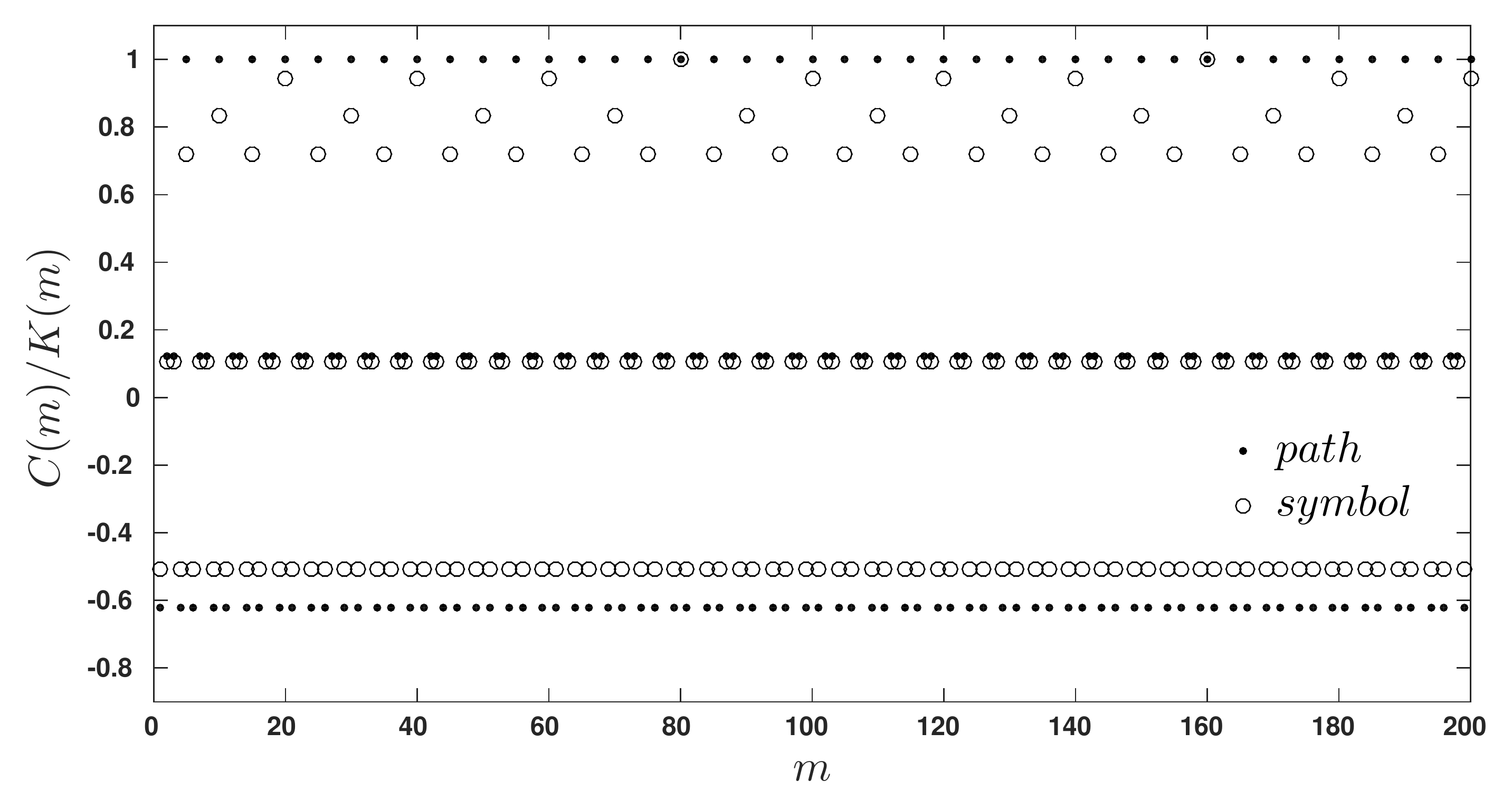}
                \vspace{-0.2cm}
                \caption{Shown are the symbol-to-symbol correlation function and the correlation function of the
                trajectory
                for the logistic map with control parameter value $r =3.7430$ (cycle $5\cdot{2^\infty }(a)$) and initial condition
                ${x_0} = 0.5$. The first ${10^5}$ iterations
                have been eliminated from our statistics in order to exclude transients, and the subsequent $n={10^8}$ iterations have been taken
                into account for the numerical calculations.}
\end{figure}
In \cite{Karam03}, on the grounds of the
Metropolis-Stein-Stein algorithm, we have established a new theorem,
namely that
\begin{equation}
 \label{eq:13}{{K_{un}}(m) = \left\{ {\begin{array}{*{20}{c}}
{ - \frac{1}{9},}&{m = 1 + 2 \cdot k}&{(k = 0,1,2,...)}\\
{\frac{1}{{18}},}&{m = 2 + 4 \cdot k}&{(k = 0,1,2,...)}\\
{\frac{{10}}{{72}},}&{m = 4 + 8 \cdot k}&{(k = 0,1,2,...)}\\
{\frac{{13}}{{72}},}&{m = 8 + 16 \cdot k}&{(k = 0,1,2,...)}\\
{\frac{{29}}{{144}},}&{m = 16 + 32 \cdot k}&{(k = 0,1,2,...)}
\end{array}} \right.}
\end{equation}
As it seems, these are the first few numerically obtained
coefficients $A_{l}$ of eq.(9). We can extend this procedure to
infinity, i.e. $m\to \infty$, and the above constructive scheme guarantees that this deep
algorithmic structure is kept in all scales. From this infinite
stratification, the infinite memory of the system at the Feigenbaum
point is revealed, as this scheme never ends.

In the following, we proceed by extending our numerical
experimentation to the structure of the normalized symbol-to-symbol
correlation function taking again into account the first $n={10^8}$
iterations of the logistic map (this scheme is depicted in Figure 2)
with initial condition $x_{0}=0.5$. The corresponding functional structure now
reads
\begin{equation}
\label{eq:14}K(m) = \left\{ {\begin{array}{*{20}{c}}
  {-\frac{1}{2},}&{m = 1 + 2 \cdot k}&{(k = 0,1,2,...)} \\
  {\frac{1}{4},}&{m = 2 + 4 \cdot k}&{(k = 0,1,2,...)} \\
  {\frac{1}{{1.6}},}&{m = 4 + 8 \cdot k}&{(k = 0,1,2,...)} \\
  {\frac{1}{{1.23}},}&{m = 8 + 16 \cdot k}&{(k = 0,1,2,...)} \\
  {\frac{1}{{1.10}},}&{m = 16 + 32 \cdot k}&{(k = 0,1,2,...)} \\
  {\sim\frac{1}{{1.049}},}&{m = 32 + 64 \cdot k}&{(k = 0,1,2,...)} \\
  {\sim\frac{1}{{1.024}},}&{m = 64 + 128 \cdot k}&{(k = 0,1,2,...)} \\
  {\sim\frac{1}{{1.012}},}&{m = 128 + 256 \cdot k}&{(k = 0,1,2,...)} \\
  {\sim\frac{1}{{1.006}},}&{m = 256 + 512 \cdot k}&{(k = 0,1,2,...)} \\
  {\sim\frac{1}{{1.002}},}&{m = 512 + 1024 \cdot k}&{(k = 0,1,2,...)} \\
  {}&{}&{}
\end{array}} \right.
\end{equation}
which has already been established in \cite{Karam03}. Moreover,
proceeding along the same lines one can calculate the
correlation function of the trajectory as it is defined in eqs.(5,6).
In this manner, it can be easily confirmed numerically that its
structure has the following simple form
\begin{equation}
\label{eq:15}C(m) = \left\{ {\begin{array}{*{20}{c}}
  { - \frac{1}{{1.0976}},}&{m = 1 + 2 \cdot k}&{(k = 0,1,2,...)} \\
  {\frac{1}{{1.2122}},}&{m = 2 + 4 \cdot k}&{(k = 0,1,2,...)} \\
  {\frac{1}{{1.0171}},}&{m = 4 + 8 \cdot k}&{(k = 0,1,2,...)} \\
  {\frac{1}{{1.0016}},}&{m = 8 + 16 \cdot k}&{(k = 0,1,2,...)} \\
  {\frac{1}{{1.00015}},}&{m = 16 + 32 \cdot k}&{(k = 0,1,2,...)} \\
  {\frac{1}{{1.00001}},}&{m = 32 + 64 \cdot k}&{(k = 0,1,2,...)} \\
  {\frac{1}{{1.000001}},}&{m = 64 + 128 \cdot k}&{(k = 0,1,2,...)} \\
  {\frac{1}{{1.0000001}},}&{m = 128 + 256 \cdot k}&{(k = 0,1,2,...)} \\
  {\frac{1}{{1.00000001}},}&{m = 256 + 512 \cdot k}&{(k = 0,1,2,...)}
\end{array}} \right.
\end{equation}
As it can be observed it possesses the known functional form of
eq.(9) apart from the numerical constants $A'_{l}$, whereas
$A'_{l}\to 1$ as $m$ becomes large. Therefore, up to now we have
established a general form for the correlation function of the
trajectory exactly at the $FP$ point supported from numerical
calculations. In the following, we shall present some arguments in
order to show that the correlation functions of the trajectory and
that obtained from the symbolic sequence possesses the same time
scales. This discussion will lead us to a conceptual understanding
for the reason that the correlation function of the trajectory (as
it is defined in eq.(5)) contains the same time scales with the
symbol-to-symbol correlation function (see eq.(7)), that is
\begin{equation}
\label{eq:17}{C_{un}(m) = {A'_l} \cdot {\delta _{m,{2^{l - 1}} \cdot
(1 + 2 \cdot k)}}}.
\end{equation}
Here for a fixed $l \in \{ 1,2,3,....\}$ the coefficients $A'_{l}$
are constants, while $k$ takes all the values from the set of
natural numbers $\{0,1,2,...\}$. Note also that the properties of a single 
position of the trajectory have already been studied in another context \cite{Robledo1,Robledo2,Robledo3}, 
where it has been shown that the trajectory exhibits the above mentioned iteration-times property. 
In the following, we would like to
show that when the symbol-to-symbol correlation function saturates, 
the correlation function of the path is saturated, too, and when the
symbol-to-symbol correlation function diminishes, the correlation
function of the path diminishes, too. Towards this direction, let us
consider the deviation from the critical value $x_{cr}=0.5$ as $0 <
{\varepsilon _i} < 0.5$ with ${\varepsilon _i}{\varepsilon _j} <
0.25$, which remains valid for all cases. As a consequence of the
above proposition we can distinguish the following four different
cases: a) A contribution in the symbolic correlation function by two
terms ${x_k} > 0.5$ (so $y_{k}=1$) and ${x_{k+l}} > 0.5$
($y_{k+l}=1$) is +1 and gives a corresponding contribution to the
trajectory correlation function of the order of $(0.5 +
\varepsilon_{i} )(0.5 + \varepsilon_{j} ) \simeq 0.25 + \varepsilon
+ \mathcal{O} ({\varepsilon ^2})$, that is more than 0.25. b)
Secondly, a contribution in the symbolic correlation function by the
terms ${x_k} < 0.5$ and ${x_{k+l}} < 0.5$ is 0 and gives a
corresponding contribution to the correlation function of the
trajectory of the order of $(0.5 - \varepsilon_{i} )(0.5 -
\varepsilon_{j} ) \simeq 0.25 - \varepsilon + \mathcal{O}
({\varepsilon ^2})$, that is less than 0.25. c) On the other hand, a
contribution in the symbolic correlation function by two terms
${x_k} > 0.5$ and ${x_{k+l}} < 0.5$ is 0 and gives a corresponding
contribution to the trajectory correlation function of the order of
$(0.5 + \varepsilon_{i} )(0.5 - \varepsilon_{j} ) \simeq 0.25 +
\mathcal{O} ({\varepsilon ^2})$, i.e. of the order of 0.25. d)
Finally, a contribution in the symbolic correlation function by the
terms ${x_k} < 0.5$ and ${x_{k+l}} > 0.5$ is 0 and gives a
corresponding contribution to the trajectory correlation function of
the order $(0.5 - \varepsilon_{i} )(0.5 + \varepsilon_{j} ) \simeq
0.25 + \mathcal{O} ({\varepsilon ^2})$, that is of the order of
0.25.

In order to clarify the meaning of the above approximations let us
consider a specific example with $x_{k}=0.7$ and $x_{k+l}=0.9$.
Then, we have ${\varepsilon _1} = 0.2$ and ${\varepsilon _2} = 0.4$.
So, the contribution to the symbolic correlation function is 1 and
as a consequence the contribution to the real correlation function
is $0.55=0.25+\epsilon>0.25$. Thus, from the above it is clear that
using such a simple argument one can predict correctly the
functional form of the correlation function of the trajectory from
the symbolic one.
\begin{table*}[ht]
\caption{In the second column of the table below the values of the
control parameter which correspond to the different accumulation
points (first column) of the logistic map are presented. In the
third and fourth columns we show the Lyapunov exponent obtained from
$n={10^8}$ iterations, including and excluding transients
respectively. We observe no significant differences due to the
augmented statistics.} \centering

\begin{tabular}{c c c c }
\hline\hline Accumulation cycle  & Accumulation point & Lyapunov
exponent& Lyapunov exponent \\
[0.5ex]
\hline ${2^\infty }$ & FP & $- {5}.{934}\cdot {10^{-5}} $ & $- {5}.{934}\cdot {10^{-5}} $ \\
\hline $3\cdot{2^\infty }$ &3.8495  &$0.0237$ &$0.0237$ \\
\hline $4 \cdot {2^\infty }$& 3.9612 & $0.0122$ &$0.0122$ \\
\hline $5 \cdot {2^\infty }(a)$&3.7430 & $-0.0021$ & $-0.0021$ \\
\hline $5 \cdot {2^\infty }(b)$&3.9065 & $0.0414$& $0.0414$ \\
\hline $5 \cdot {2^\infty }(c)$& 3.99032& $-0.0039$  &$-0.0039$ \\
\hline $6 \cdot{2^\infty }(a)$&3.6327& $-0.0073$  & $-0.0073$ \\
\hline $6 \cdot {2^\infty }(b)$  & 3.937649 &$0.0127$  & $0.0127$ \\
\hline $6 \cdot {2^\infty }(c)$ & 3.977800 & $0.0077$ &$0.0077$ \\
\hline $6 \cdot {2^\infty }(d)$ &3.997586 & $-0.0133$ & $-0.0133$ \\
[1ex] \hline \hline
\end{tabular}
\label{table:nonlin}
\end{table*}

As the structure of the correlation function of the trajectory for
the $2^{\infty}$ scenario has been fairly understood, let us proceed
with the next accumulation points. Thus, we further consider the $3
\cdot {2^\infty }$ scenario which corresponds to the control
parameter value $r=3.8495$ of the logistic map (see Table I). In
this manner, one can evaluate the symbol-to-symbol correlation for
this scenario ($3\cdot{2^\infty}$), using the same numerical
procedure and statistical sample as previously. From this
calculation we can conclude that the normalized symbol-to-symbol
correlation function for the $3\cdot{2^\infty}$ accumulation point
has the following form
\begin{equation}
 \label{eq:16}C(\tau ) = {A_l} \cdot {\delta _{\tau ,3 \cdot {2^l}(1 + 2k)}} + {B_1} \cdot {\delta _{\tau ,1 + 3 \cdot k}} +
 {B_2} \cdot {\delta _{\tau ,2 + 3 \cdot k}},
\end{equation}
where for a given $l \in \{ 1,2,3,....\}$, $A_{l}$, $B_{1}$, $B_{2}$
are constants and $k$ takes all the values from the set of natural
numbers $\{0,1,2,...\}$.

\begin{table*}[ht]
\caption{In the table below we present the first few coefficients of
eq.(14) for the normalized symbol-to-symbol correlation function at
each accumulation point. The corresponding mean value for the first
$n={10^8}$ iterations of the logistic map is also provided.}
\centering

\begin{tabular}{c c c c c c c c c c}
\hline\hline Scenario & mean value & ${A_0}$ & ${A_1}$ & ${A_2}$ &
${B_1}$
& ${B_2}$  & ${B_3}$ & ${B_4}$ & ${B_5}$  \\
\hline ${2^\infty }$ & $0.6666$ & $-\frac{1}{2}$ & $\frac{1}{4}$ & $
\frac{1}{{1.6}}$ & $-$ & $-$
&$-$ & $-$ & $-$\\
\hline $3 \cdot {2^\infty }$  & $0.4381$ & $\frac{1}{{1.7405}}$ &
$\frac{1}{{1.3362}}$ & $\frac{1}{{1.0952}}$ & $-\frac{1}{{2.8240}}$
&
$-\frac{1}{{2.8240}}$ & $-$ & $-$ & $-$ \\
\hline $4 \cdot {2^\infty }$  & $0.3342$ & $\frac{1}{{1.6083}}$ &
$\frac{1}{{1.2249}}$ & $\frac{1}{{1.1078}}$ & $ -
\frac{1}{{8.089362}}$ &
$ - \frac{1}{2}$ & $ - \frac{1}{{8.089365}}$ & $-$ & $-$ \\
\hline $5 \cdot {2^\infty }(a)$ & $0.6625$ & $\frac{1}{{1.3880}}$ &
$\frac{1}{{1.2015}}$ & $\frac{1}{{1.0592}}$ & $ -
\frac{1}{{1.9630}}$ & $\frac{1}{{9.4768}}$ & $\frac{1}{{9.4768}}$ &
$-
\frac{1}{{1.9630}}$ & $-$ \\
\hline $5 \cdot {2^\infty }(b)$ & $0.5420$ & $\frac{1}{{1.3050}}$ &
$\frac{1}{{1.1496}}$ & $\frac{1}{{1.1139}}$ & $ -
\frac{1}{{25.4566}}$ &
$ - \frac{1}{{2.6470}}$ & $ - \frac{1}{{2.6470}}$ & $ - \frac{1}{{25.4566}}$ & $-$ \\
\hline $5 \cdot {2^\infty }(c)$ & $0.2500$ & $\frac{1}{{1.3636}}$ &
$\frac{1}{{1.3636}}$ & $1$ & $ - \frac{1}{{15}}$ &
$ - \frac{1}{3}$ & $ - \frac{1}{3}$ & $- \frac{1}{{15}}$ & $-$ \\
\hline $6 \cdot {2^\infty }(a)$ & $0.7708$ & $\frac{1}{{1.5475}}$ &
$\frac{1}{{1.1337}}$ & $\frac{1}{{1.1337}}$ & $ -
\frac{1}{{3.3636}}$ & $\frac{1}{{17.6957}}$ & $- \frac{1}{{3.3636}}$
& $\frac{1}{{17.6957}}$ & $-
\frac{1}{{3.3636}}$ \\
\hline $6 \cdot {2^\infty }(b)$ & $0.5529$ & $\frac{1}{{1.2725}}$ &
$\frac{1}{{1.1402}}$ & $\frac{1}{{1.0478}}$ & $-\frac{1}{{7.4466}}$
&
$-\frac{1}{{2.8699}}$ & $\frac{1}{{8.9594}}$ & $-\frac{1}{{2.8699}}$ & $-\frac{1}{{7.4466}}$ \\
\hline $6 \cdot {2^\infty }(c)$ & $0.4468$ & $\frac{1}{{1.2744}}$ &
$\frac{1}{{1.1387}}$ & $\frac{1}{{1.049}}$ & $\frac{1}{{9.0602}}$ &
$-\frac{1}{{7.5024}}$ & $-\frac{1}{{1.2382}}$ & $-\frac{1}{{7.5024}}$ & $\frac{1}{{9.0602}}$ \\
[0.5ex] \hline $6 \cdot {2^\infty }(d)$ & $0.2500$ &
$\frac{1}{{1.8}}$ & $1$ & $1$
& $\frac{1}{9}$ & $-\frac{1}{3}$ & $-\frac{1}{3}$ & $-\frac{1}{3}$ & $\frac{1}{9}$ \\[1ex] \hline \hline
\end{tabular}
\label{table:nonlin}
\end{table*}

\begin{table*}[ht]
\caption{In the table below we present the first few coefficients of
eq.(14) for the normalized correlation function of the trajectory at
each accumulation point. The corresponding mean value for the first
$n={10^8}$ iterations of the logistic map is also provided.}
\centering

\begin{tabular}{c c c c c c c c c c}
\hline\hline Scenario & mean value & ${A_0}$ & ${A_1}$ & ${A_2}$ &
${B_1}$
& ${B_2}$ & ${B_3}$ & ${B_4}$ & ${B_5}$  \\
\hline ${2^\infty }$ & $0.6476$ & $-\frac{1}{{1.0976}}$ &
$\frac{1}{{1.2122}}$ & $\frac{1}{{1.0171}}$ & $-$ & $-$ &
$-$ & $-$ & $-$ \\
\hline $3 \cdot {2^\infty }$ & $0.5313$  & $\frac{1}{{1.0029}}$ &
$\frac{1}{{1.0028}}$ & $\frac{1}{{1.00002}}$ & $-\frac{1}{{2.0053}}$
&
$-\frac{1}{{2.0009}}$ & $-$ & $-$ & $-$ \\
\hline $4 \cdot {2^\infty }$ & $0.4181$  & $\frac{1}{{1.0001}}$ &
$\frac{1}{{1.00001}}$ & $\frac{1}{{1.000001}}$ &
$-\frac{1}{{6.0840}}$ &
$-\frac{1}{{1.4899}}$ & $-\frac{1}{{6.0827}}$ & $-$ & $-$ \\
\hline $5 \cdot {2^\infty }(a)$  & $0.6311$ & $\frac{1}{{1.0008}}$ &
$\frac{1}{{1.0007}}$ & $\frac{1}{{1.00001}}$ & $-\frac{1}{{1.6112}}$
&
$\frac{1}{{8.2817}}$ & $\frac{1}{{8.2847}}$ & $-\frac{1}{{1.6119}}$ & $-$ \\
\hline $5 \cdot {2^\infty }(b)$  & $0.5472$ & $\frac{1}{{1.00008}}$
& $\frac{1}{{1.00001}}$ & $\frac{1}{{1.000009}}$ & $-
\frac{1}{{2.7923}}$ &
$ - \frac{1}{{7.0485}}$ & $- \frac{1}{{7.0489}}$ & $- \frac{1}{{2.7925}}$ & $-$ \\
\hline $5 \cdot {2^\infty }(c)$  & $0.3382$ & $\frac{1}{{1.000004}}$
& $\frac{1}{{1.0000004}}$ & $1$ & $\frac{1}{{47.222}}$ &
$-\frac{1}{{1.918743}}$ & $-\frac{1}{{1.918742}}$ & $\frac{1}{{47.2256}}$ & $-$ \\
\hline $6 \cdot {2^\infty }(a)$  & $0.6609$ &
$\frac{1}{{{{1}}{{.00109}}}}$ & $\frac{1}{{1.000098}}$ &
$\frac{1}{{1.0000088}}$ & $ - \frac{1}{{{{1}}{{.29384}}}}$ &
$\frac{1}{{{{1}}{{.67651}}}}$ & $-\frac{1}{{{{1}}{{.54632}}}}$ & $\frac{1}{{{{1}}{{.67777}}}}$ & $- \frac{1}{{{{1}}{{.29473}}}}$ \\
\hline $6 \cdot {2^\infty }(b)$  & $0.5504$ & $\frac{1}{{1.00001}}$
& $\frac{1}{{1.0000001}}$ & $\frac{1}{{1.00000008}}$ & $-
\frac{1}{{{{6}}{{.87663}}}}$ &
$- \frac{1}{{{{4}}{{.12738}}}}$ & $- \frac{1}{{{{4}}{{.4526}}}}$ & $- \frac{1}{{{{4}}{{.12735}}}}$ & $- \frac{1}{{{{6}}{{.8768}}}}$ \\
\hline $6 \cdot {2^\infty }(c)$  & $0.4610$ & $\frac{1}{{1.000002}}$
& $\frac{1}{{1.0000002}}$ & $\frac{1}{{1.00000002}}$ & $-
\frac{1}{{{{22}}{{.6942}}}}$ &
$- \frac{1}{{{{8}}{{.28704}}}}$ & $- \frac{1}{{{{1}}{{.49136}}}}$ & $- \frac{1}{{{{8}}{{.28707}}}}$ & $- \frac{1}{{{{22}}{{.6945}}}}$ \\
[0.5ex] \hline $6 \cdot {2^\infty }(d)$  & $0.2826$ &
$\frac{1}{{1.0000002}}$ & $1$ & $1$
& $\frac{1}{{{{7}}{{.78286}}}}$ & $- \frac{1}{{{{2}}{{.6275408}}}}$ & $- \frac{1}{{{{2}}{{.01691}}}}$ & $- \frac{1}{{2.6275407}}$ & $\frac{1}{{{{7}}{{.78286}}}}$ \\[1ex] \hline \hline
\end{tabular}
\label{table:nonlin}
\end{table*}
The first few numerical values of the above coefficients for the
correlation function of the symbolic sequence are presented in Table
II (see second row). As one can easily verify after a
straightforward numerical computation the same structure is observed
for the correlation function of the trajectory, apart from numerical
constants ${A'_l}$, ${B'_1}$, ${B'_2}$ which depend on the detailed
form of the map (see Table III, third row). In the same manner, one
can perform the same calculations for the higher accumulation points
and find a similar structure for both the correlation function of
the trajectory and that obtained from the symbolic sequence. Indeed,
in Tables II and III we extend these considerations to the $4 \cdot
{2^\infty }$, $5 \cdot {2^\infty }(a)$, $5 \cdot {2^\infty }(b)$, $5
\cdot {2^\infty }(c)$, $6 \cdot {2^\infty }(a)$, $6 \cdot {2^\infty
}(b)$, $6 \cdot {2^\infty }(b)$, $6 \cdot {2^\infty }(c)$ and $6
\cdot {2^\infty }(d)$ (see also Figures 3,4) accumulation points and
we present the corresponding coefficients that may appear in the
respective correlation function for each cycle. Notice that the
relevant information about the accumulations points and the
corresponding patterns have been found in \cite{Freud96}. As for the
lower cycles both the correlation function of the trajectory and the
symbolic one posses the same time scales. To illustrate the above,
Figure 3 shows the symbol-to-symbol versus the correlation function
of the trajectory for the $3 \cdot {2^\infty }$ cycle. We observe
that the correlation function for the trajectory is always larger
from the symbolic one. The latter can also be confirmed by a direct
comparison of the coefficients $A_{i}$ or $B_{i}$ among the
different types of correlations presented in Tables II and III. The
previous observations also hold for higher accumulation points, e.g.
the $5\cdot {2^\infty }(a)$ cycle as shown in Figure 4.
Additionally, from the Tables II, III it is clearly seen that for
the same type of correlations the coefficients $A_{i}$ are
increasing for higher accumulation points while some of the
constants ${B_i}$ may coincide. This is more rare for the
correlation function of the trajectory.

As a result of the above studies we can conclude that the suggested
general form for the symbol-to-symbol correlation function at the $FP$ point of the $m \cdot
{2^\infty }$ attractor reads
\begin{equation}
 \label{eq:17}C(\tau ) = {A_l} \cdot {\delta _{\tau ,m \cdot {2^l}(1 + 2k)}} + {B_1} \cdot {\delta _{\tau ,1 + m \cdot k}} +
 {B_2} \cdot {\delta _{\tau ,2 + m \cdot k}} + ... + {B_{(m - 1)}} \cdot {\delta _{\tau ,(m - 1) + m \cdot
 k}},
\end{equation}
where as usual for a given $l \in \{ 1,2,3,....\}$ the coefficients
$A_{l}$, $B_{1}$, $B_{2}$,...,$B_{m-1}$ are constants depending only
on $l$, $m$ denotes the number of the accumulation point that we
consider and $k$ takes all the values from the set of natural
numbers $\{0,1,2,...\}$. The corresponding form for the correlation
function of the trajectory remains the same apart from numerical
constants ${A'_l}$, ${B'_1}$, ${B'_2}$,... which as it has been
mentioned previously they depend on the detailed form of the map.

 \section{Summary and Conclusions }

The correlation function is an important quantity measuring
correlations in many branches of physics. Obviously, there are also
other interesting quantities as for instance the (conditional)
block-entropies, the transinformation, the Kolmogorov-Sinai entropy
etc. However, it does provide an important measure of correlations
by itself.

In the present paper the correlation function of the
trajectory at the Feigenbaum point is numerically investigated with
careful numerical experimentation. Comparing with the
symbol-to-symbol correlation function discussed in the literature
theoretically and numerically we observe that it contains the same
time scales, that is, it has the same functional form. This result
has been also justified up to a good approximation by presenting
simple arguments. Moreover, we have generalized these results for
the case of an arbitrary $m \cdot 2^{\infty}$ Feigenbaum non chaotic
multifractal attractor. Finally, we have arrived to an empirical
formula summarizing the results.

 To recapitulate, we are in position to justify the analytical form of the correlation
function of the trajectory from first principles (the MSS algorithm)
and in a systematic way, apart from numerical constants which depend
on the detailed functional form of the map. Apart from their
mathematical beauty such ideas find important practical applications
ranging from precursory signals \cite{Karam} to DNA sequence
analysis \cite{Athan,Karam2}, Heart beat rhythms \cite{Karam1} and
Linguistics Processes. In this manner, it is still an open problem
what information one can extract by using such complexity measures
in real experimental time-series and the physical explanation of the
correlation function of the trajectory and that obtained from the
symbolic sequence for a specific problem. A second path towards this 
direction would be  
the generalization of the form of the correlation function for more 
complex maps as well as non-unimodal maps, see for instance \cite{Grebogi}.

{}

\end{document}